\documentstyle[PASJadd]{PASJ95}
\begin{document}

\title{Distribution and Kinematics of Molecular Gas in Barred Spiral Galaxies.\\
 I. NGC 3504}

\author{Nario {\sc Kuno}, Kohta {\sc Nishiyama}, Naomasa {\sc Nakai}, Kazuo {\sc Sorai}, 
Baltasar {\sc Vila}-{\sc Vilar\'o}\thanks{Present address: The Submillimeter Telescope Observatory, Steward Observatory, University of Arizona, Tucson, AZ 85721, USA}\\
{\it Nobeyama Radio Observatory\thanks{Nobeyama Radio Observatory (NRO) is a branch of the National Astronomical Observatory, an inter-unviersity research institute operated by the Ministry of Education, Science, Sports and Culture.}, Minamimaki-mura, Minamisaku-gun, Nagano 384-1305}\\
{\it E-mail(NK): kuno@nro.nao.ac.jp}\\
and\\
Toshihiro {\sc Handa}\\ 
{\it Institute of Astronomy, The University of Tokyo,
2-21-1 Osawa, Mitaka, Tokyo 181-0015}}

\abst{We present the results of CO mapping observations of the barred spiral galaxy NGC 3504 with the Nobeyama 45-m telescope. The distribution of the molecular gas shows offset ridges which correspond to the distribution of H\,{\footnotesize II} regions along the bar. The velocity perpendicular to the bar decreases abruptly at the ridge. The velocity change implies that the molecular gas changes the direction of its motion to parallel to the bar at the ridge. Since the position angle of the major axis of the bar and the line of nodes are almost the same in NGC 3504, an upper limit to the pattern speed of the bar can be derived directly from the radial velocity. The resultant upper limit is 41 km s$^{-1}$ kpc$^{-1}$, which is much smaller than that derived based on an assumption that the corotation radius is located at the end of the bar (77 km s$^{-1}$ kpc$^{-1}$). The corotation radius derived from our upper limit is more than two-times larger than the length of the semi-major axis of the bar in NGC 3504.}

\kword{galaxies: bar --- galaxies: individual (NGC 3504) --- galaxies: 
structure --- ISM: molecules}

\maketitle

\section{Introduction}

Many numerical simulations have been performed up until the present to study the distribution and dynamics of the interstellar medium in barred spiral galaxies. The behavior of the interstellar medium in a bar potential and the influence of a bar on the evolution of galaxies have been evaluated in those simulations (e.g., Athanassoula 1992; Combes, Gerin 1985). However, observational data are still insufficient to compare with the numerical simulation results. Since it takes so much time to map a galaxy with sufficient high angular resolution to resolve the bar and spiral structures, only a few barred spiral galaxies have been fully mapped in molecular lines so far (e.g., Sempere, Garcia-Burillo 1997).  

We conducted CO mapping observations of barred spiral galaxies with the 45-m telescope at the Nobeyama Radio Observatory (NRO). The combination of the telescope and the $2\times2$-beam SIS focal plane array receiver (Sunada et al.\ 1995) results in a high angular resolution and high mapping performance, which allows highly efficient mapping observations of external galaxies. The purpose of our project is to investigate the distribution and kinematics of molecular gas in barred spiral galaxies. As a sample, we selected six galaxies (M 83, NGC 4321, NGC 253, NGC 3504, NGC 4303, NGC 6951) which are nearby and bright enough in CO to map and resolve the large-scale structures, such as the bar and spiral arms with the NRO 45-m telescope.   

In this paper we present the results of one of our targets, NGC 3504. The basic parameters of NGC 3504 are listed in table 1. The central region of this galaxy has been previously mapped in CO by both an interferometer (OVRO: Kenney et al. 1993) and a single-dish telescope (IRAM 30m: Planesas et al.\ 1997). While these maps show only a concentration of molecular gas in the center of the galaxy, we could obtain the distribution of molecular gas in the whole of the bar region because our observations were more sensitive and covered a larger area. From these data we were able to derive an upper limit to the pattern speed of the bar using our original method and the locations of the resonances. It is very important to know the pattern speed and locations of resonances in barred spiral galaxies, since it has been suggested that the resonances are related to large-scale structures in galaxies, such as inner rings, outer rings, and bar lengths (e.g., Combes 1996). Moreover, since the behavior of interstellar gas in a bar potential is different for fast and slow bars, the difference in the pattern speed of the bar affects the evolution of galaxies (Combes, Elmegreen 1993). In previous studies, the pattern speed of a bar was derived with an assumption that some feature is located at a resonance except for a few cases (Canzian 1993; Tremaine, Weinberg 1984; Sempere et al. 1995). In this framework, however, discussion about the relations between resonances and galactic structures is nonsense, since some results are settled in the assumption. Without such an assumption, we obtained a strict upper limit of the pattern speed in NGC 3504. It thus becomes possible to address the relation between the pattern speed and galactic structures. 


\begin{table*}[t]
\begin{center}
Table~1.\hspace{4pt}Parameters of NGC 3504.
\end{center}
\vspace{6pt}
\begin{tabular*}{\textwidth}{@{\hspace{\tabcolsep}
\extracolsep{\fill}}p{20pc}c}
\hline\hline\\[-6pt]
Center position (1950.0)$^\ast$ & R.A.\ = $11^{\rm h}00^{\rm m}28^{\rm s}\hspace{-5pt}.\hspace{2pt}53$\\
                                      & Decl.\ = $28^{\circ}14'31''\hspace{-5pt}.\hspace{2pt}2$\\
Morphological type$^\dagger$ & ($R_{1}$)SAB($rs$)ab\\
Distance$^\ddagger$ & 20 Mpc\\
Systemic velocity (LSR)$^\ddagger$ & 1535 km s$^{-1}$\\
Position angle of major axis$^\ddagger$ & $149^{\circ}$\\
Position angle of bar$^\ddagger$ & $143^{\circ}$\\
Inclination angle$^{\S}$ & $22^{\circ}$\\
\hline
\end{tabular*}

\vspace{6pt}\par\noindent
$*$ Condon et al.\ (1990)
\par\noindent
$\dagger$ de Vaucouleurs et al.\ (1991)
\par\noindent
$\ddagger$ Kenney et al.\ (1993)
\par\noindent
$\S$ Grosb{\o}l (1985)
\end{table*}

\section{Observations}

Observations of $^{12}$CO ($J$ = 1--0) emission (rest frequency 115.271204 GHz) were made between 1995 December and 1997 January with the 45-m telescope at Nobeyama Radio Observatory. The full half-power beam width (HPBW) at 115 GHz was $16''$, which corresponds to 1.6 kpc at the distance of NGC 3504 (20 Mpc). The aperture and main-beam efficiencies were $\eta_{\rm a} = 0.38$ and $\eta_{\rm mb} = 0.50$, respectively. 

A $2\times2$-beam SIS focal-plane array was used as the receiver frontend. With this receiver, we could observe simultaneously four different positions separated on the sky by $34''$ each. 2048-channel wide-band acousto-optical spectrometers (AOS) were used as receiver backends. The frequency resolution and channel spacing were 250 kHz and 125 kHz, respectively, providing a total bandwidth of 250 MHz.  At 115 GHz the corresponding velocity resolution and velocity coverage are 0.65 km s$^{-1}$ and 650 km s$^{-1}$, respectively. Calibration of the line intensity was made using the chopper-wheel method (Ulich, Haas 1976), yielding the antenna temperature ($T_{\rm A}^{*}$) corrected for both atmospheric and antenna ohmic losses. Since the sideband ratios of the four beams of the array receiver were different, we calibrated the intensity by observing of IRC +10216 in CO both with the array receiver and a single beam SIS receiver with an SSB (single sideband) filter. The system noise temperatures, including the atmospheric effect and the antenna ohmic loss, were in the range of 550 -- 1000 K (SSB) during the observations. In this paper, we use the main beam brightness temperature scale, $T_{\rm MB}  \equiv T_{\rm A}^{*}/\eta_{\rm mb}$, and we assume that $T_{\rm MB}$ is equal to the brightness temperature.

We observed 72 points in a region of $66''\times132''$ with a grid spacing of $11''$ parallel and perpendicular to the major axis of the galaxy at a position angle of $149^{\circ}$. The telescope pointing was checked and corrected every hour by observing the SiO maser emission of the late type star R Leo at 43 GHz. The absolute pointing accuracy was better than $5''$ (peak value) throughout the observations.

\section{Molecular Gas Distribution}

\subsection{Molecular Gas Ridge}
Figure 1 shows a profile map of the CO emission and figure 2 shows the integrated intensity ($I_{\rm CO} \equiv \int T_{\rm MB} dv$ [K km s$^{-1}$]) map of NGC 3504. The map shows the presence of molecular gas ridges along the leading edges of the bar (figure 3), while only the central peak was found in  previous CO observations (Kenney et al. 1993; Planesas et al. 1997). In NGC 3504, there are many H\,{\footnotesize II} regions along the bar, as can be seen in the H$\alpha$ image in Kenney et al.\ (1993). These H\,{\footnotesize II} regions seem to lie along the molecular-gas ridges (figure 4).


\begin{fv}{1}
{0cm}
{Profile map of the CO emission in NGC 3504. $X$ and $Y$ are parallel to the minor and major (P.A.\ = $149^{\circ}$) axes, respectively. The center position, ($X$, $Y$) = (0$''$, 0$''$), is ($\alpha_{1950}$, $\delta_{1950}$) = ($11^{\rm h}00^{\rm m}28^{\rm s}\hspace{-5pt}.\hspace{2pt}53$, $28^{\circ}14'31''\hspace{-5pt}.\hspace{2pt}2$).}
\end{fv}


\begin{fv}{2}
{0cm}
{CO integrated intensity map of NGC 3504. $X$ and $Y$ are parallel to the minor and major axes, respectively. The contour levels are 2, 4, 6, 10, 20, 40, 60 K km s$^{-1}$.}
\end{fv}


\begin{fv}{3}
{0cm}
{CO integrated intensity map (contours) overlaid on optical images (Sandage 1961). The contour levels are 2, 4, 6, 8, 10, 15, 20, 30, 40, 50, 60 K km s$^{-1}$. North is upside and East is leftside. The brightness and contrast of the optical images were controled to see the structure inside of the bar (right) and large scale structure of the bar and the arms (left).}
\end{fv}


\begin{fv}{4}
{0cm}
{CO integrated intensity map (contours) overlaid on H$\alpha$ image from Kenney et al.\ (1993). The contour levels are the same as figure 3. North is upside and East is leftside.}
\end{fv}

As can be clearly seen in figure 2, most of the molecular gas is concentrated in the central region of this galaxy. At the center [($X$, $Y$) = ($0''$, $0''$)] the surface density corrected for the inclination of the galaxy is 222 $\MO$ pc$^{-2}$ assuming a galactic conversion factor of $N$(H$_{2}$)/$I_{\rm CO}$ = $2.3\times10^{20}$ cm$^{-2}$ [K km s$^{-1}]^{-1}$ (Strong et al.\ 1988). On the other hand, the average surface density of 8 points [($X$, $Y$) = ($-11''$, $22''$),  ($-11''$, $33''$),  ($0''$, $22''$), ($0''$,$33''$), ($0''$, $-22''$), ($0''$, $-33''$), ($11''$, $-22''$), ($11''$, $-33''$)] in the ridges is $20.5\pm$3.4 $\MO$ pc$^{-2}$. The central region is elongated along the E--W direction and the ridges begin from the edges of the elongated structure. These structures of molecular gas in the bar resemble those found in M 83, which also shows molecular-gas ridges along the leading edges of the bar and an elongated structure in the bar (Handa et al.\ 1990, 1994).

Figure 5 shows the intensity profiles along the lines of $Y = 22''$ (circles) and $-22''$ (triangles). The curves indicate the results of fitting with a Gaussian function. The widths of the ridges are fairly large: $24''$ for $Y = 22''$ and $22''$ for $Y = -22''$, respectively. The intrinsic widths of the ridge corrected for our beam size ($16''$) are $18''$ and $15''$, which correspond to 1.7 and 1.5 kpc, respectively. It should be noted, however, that these widths are upper limits, since our sampling interval (11$''$) is larger than that for Nyquist sampling. Although the molecular ridges connect to the spiral arms, the molecular arms are not prominent.  


\begin{fv}{5}
{0cm}
{CO intensity profiles along the lines of $Y = 22''$ (circles) and $-22''$ (triangles). The curves show the results of Gaussian fitting.}
\end{fv}

\subsection{Radial Distribution}
Figure 6 shows the radial distribution of the surface density of the molecular gas corrected for the inclination of the galactic disk. The surface density decreases monotonically with galactocentric distance. If we fit the radial distribution with an exponential function [$\sigma$ = $\sigma_{0}\exp{(-R/R_{\rm e})}$], the scale length, $R_{\rm e}$, is $13''$, which corresponds to 1.2 kpc at the distance of the galaxy. The total mass of the molecular gas within our mapping area is $1.4\times10^{9}$ $\MO$ which is 1\% of the dynamical mass within 6 kpc ($1\times10^{11}$ $\MO$), derived by following equation assuming a spherical mass distribution:

\begin{equation}
M_{\rm dyn}(R) = \frac{RV(R)^{2}}{G},
\end {equation}
where $R$ is the galactocentric radius and $V(R)$ is the rotation velocity at $R$.

Most of the barred spiral galaxies which have been observed in CO have a central peak and condensations of molecular gas at the ends of the bar. Thus, their radial distribution of the surface density of the molecular gas usually shows a central peak and a secondary peak near the radius of the bar ends (Nakai 1992; Nishiyama, Nakai 1998). The CO map of NGC 3504 also shows condensation at the end of the bar on the southern side. The condensation of the molecular gas corresponds to complexes of H\,{\footnotesize II} regions (figure 4). However, the secondary peak at the ends of the bar can not be seen in the radial distribution. The low-resolution H\,{\footnotesize I} map in van Moorsel (1983) shows intensity peaks at the beginning of the spiral arms. This may suggest that the condensations or the secondary peak may be more apparent in the total gas distribution (H\,{\footnotesize I} + H$_{2}$).


\begin{fv}{6}
{0cm}
{Radial distribution of the surface density of molecular gas corrected for the inclination of 22$^{\circ}$. The line indicates the result of exponential fitting [$\sigma$(H$_{2}$) = 157 $\exp{(-R['']/13)}$ $\MO {\rm pc}^{-2}$]. The arrow indicates the bar length.}
\end{fv}

\section{Kinematics}

\subsection{Velocity Field and Rotation Curve}
Figure 7 is the velocity field derived from the intensity-weighted mean velocity of each spectrum. In the central region, the isovelocity contours are nearly parallel to the X axis, which coresponds to the minor axis of the galaxy. However, the isovelocity contours are apparently tilted in the ridges along the bar against the X axis, indicating non-circular motion in the bar, as mentioned in the following sections. Such disturbance of the isovelocity contours is often seen in the bar region of barred spiral galaxies (e.g., Laine et al. 1999; Sorai et al. 2000). Furthermore, large velocity gradients at the ridges can be seen in the position--velocity diagram, as shown in the next section.

Figure 8 is a position--velocity (P--V) diagram along the major axis. Kenney et al.\ (1993) showed that there is a discrepancy between the CO velocities and the H$\alpha$ velocities in the inner region ($R < 10''$). Our result is consistent with that of Kenney et al.\ (1993). On the other hand, the velocity of our CO data are consistent with H$\alpha$ for a larger radius ($R > 15''$). Although the H$\alpha$ velocities at $R > 40''$ tend to be larger than CO, our data are not sufficient to confirm the trend. The rotation curve derived by fitting the CO ($R < 10''$; Kenney et al. 1993), H$\alpha$ ($R > 15''$; Peterson 1982) and H\,{\footnotesize I} ($R = 70''.6$; van Moorsel 1983) data rises steeply at the center and has a dip at $R$ $\approx$ 20$''$. 


\begin{fv}{7}
{0cm}
{Velocity field derived from the intensity-weighted mean velocity. The contours are from 1445 km s$^{-1}$ to 1610 km s$^{-1}$ with a step of 15 km s$^{-1}$.}
\end{fv}


\begin{fv}{8}
{0cm}
{Position--Velocity diagram along the major axis. The lowest contour and contour interval are $T_{\rm MB}$ = 60 mK and 30 mK, respectively. Circles are the CO data obtained with an interferometer by Kenney et al.\ (1993). Triangles are the H$\alpha$ data from Peterson (1982). The square is the H\,{\footnotesize I} data from van Moorsel (1983). The solid line represents the rotation curve obtained by fitting the CO, H$\alpha$ and H\,{\footnotesize I} data.}
\end{fv}

\subsection{Streaming Motions along the Bar}
Figure 9 shows a P--V diagram along the line of $Y = 22''$, which is perpendicular to the bar. There is a large velocity gradient at the molecular ridge at $X = -11''$. The velocity along the line of sight is the velocity component perpendicular to the bar in NGC 3504, because the position angle of the major axis of the bar and the line of nodes are almost the same. The large velocity gradient seen in figure 9 means that the velocity component perpendicular to the bar decreases abruptly at the leading edge of the bar. The change of about 50 km s$^{-1}$ along the line of sight corresponds to about 130 km s$^{-1}$ in the galactic plane assuming an inclination angle of $22^{\circ}$. Similar velocity changes across a bar have been found in NGC 7479 (Laine et al.\ 1999). The velocity change of 130 km s$^{-1}$ in NGC 3504 is 1.5 times smaller than that in NGC 7479 (200 km s$^{-1}$).

The velocity change at the bar is consistent with many numerical simulations (e.g., Athanassoula 1992) where a shock occurs at the leading edge of the bar and gas moves along the bar. Figure 10 shows a schematic diagram of the motion of gas in a bar. Since the change seems to occur within our $16''$ beam size, we obtain a lower limit of the velocity gradient of 84 km s$^{-1}$ kpc$^{-1}$, which is comparable with that seen in the arms of the grand-design spiral M 51 (Kuno, Nakai 1997).


\begin{fv}{9}
{0cm}
{Position--velocity diagram along the $Y = 22''$. The lowest contour and contour interval are $T_{\rm MB}$ = 50 mK and 30 mK, respectively. The thick horizontal line indicates the systemic velocity of NGC 3504 ($V_{\rm LSR} = 1535$ km s$^{-1}$). The gray horizontal line indicates the velocity corresponding to the pattern speed of 77 km s$^{-1}$ kpc$^{-1}$ derived by Kenney et al.\ (1993). The arrow indicates the location of the molecular gas ridge.}
\end{fv}


\begin{fv}{10}
{0cm}
{Schematic diagram of gas motion in the bar region. The arrow indicates the orbit of a molecular cloud. The thick solid lines represent molecular gas ridges.}
\end{fv}

\subsection{Pattern Speed of the Bar}
For spiral galaxies, a rotation curve within a corotation radius gives an upper limit of the pattern speed. This is  because gases rotate faster than the pattern, such as spiral arms and bars, within the corotation radius. Thus, we can derive an upper limit of the pattern speed by observing the radial velocity along the major axis of a galaxy. The observed velocity is expressed by 
\begin{equation}
V_{\rm obs}(R) = V_{\rm sys}+(R{\Omega}_{\rm p}+V_{\theta})\sin{i},
\end{equation}
where $V_{\rm sys}$ is the systemic velocity of the galaxy, $R$ radius, ${\Omega}_{\rm p}$ pattern speed, $V_{\theta}$ rotation velocity in the rest frame of the pattern and $i$ inclination angle of the galaxy. Since $V_{\theta} \gtsim 0$ within the corotation radius,

\begin{equation}
\Omega_{\rm p} \ltsim \frac{V_{\rm obs}-V_{\rm sys}}{R\sin{i}}.
\end{equation}
Usually we can not obtain a strict upper limit, since we do not know the location of the corotation radius, and the rotation curve is often flat.
For barred spiral galaxies which have a bar parallel to the major axis of the galaxy, however, we can obtain an upper limit to the pattern speed only by using the assumption that the corotation is located beyond the end of the bar. Since material rotates in an elliptical orbit elongated along the bar, $V_{\theta}$ in equation (1) must be small in a bar. Especially, $V_{\theta}$ for interstellar gas is expected to be very small due to a shock at the bar (Athanassoula 1992; Regan et al. 1999). Therefore, we can get an upper limit which is close to the true pattern speed of the bar. In that case, the observed rotation curve looks like a rigid rotation with an angular velocity coresponding to the pattern speed of the bar. We suggest that this is the reason why galaxies which have such a configuration show a rigid-like rotation curve in the bar region (NGC 2903: Nishiyama, Nakai 1998; M 83: Handa et al. 1990; UGC 2855: H\"uttemeister et al. 1999), while the rotation curves of other barred spirals rise steeply near the center (Nishiyama, Nakai 1998).

NGC 3504 is one of such useful galaxies. Taking into acount the offset of the molecular ridge from the major axis, equation (1) and (2) are modified as follows:

\begin{equation}
V_{\rm obs}(R)=V_{\rm sys}+(R\Omega_{\rm p}\cos{\theta}+V\sin{\phi})\sin{i},
\end{equation}

\begin{equation}
\Omega_{\rm p}=\frac{V_{\rm obs}-V_{\rm sys}-V\sin{\phi}\sin{i}}{R\cos{\theta}\sin{i}},
\end{equation}
where $V$, $\phi$, and $\theta$ are defined as in figure 10. If the molecular gas rotates faster than the pattern in the bar (i.e., there is a corotation radius out of the bar), $0^{\circ} \ltsim \phi \ltsim 180^{\circ}$ and $V\sin{\phi}\sin{i} \gtsim 0$ and, thus, we can obtain an upper limit of the pattern speed as

\begin{equation}
\Omega_{\rm p} \ltsim \frac{V_{\rm obs}-V_{\rm sys}}{R\cos{\theta}\sin{i}}.
\end{equation}
Assuming that the lower velocity component $V_{\rm obs}=1560$ km s$^{-1}$ at $X = -11''$ in figure 9 represents the gas which moves along the bar after a shock at the bar, the pattern speed of the bar in NGC 3504 is $\Omega_{\rm p} \ltsim 31$ km s$^{-1}$ kpc$^{-1}$. This upper limit is smaller than the pattern speed of 77 km s$^{-1}$ kpc$^{-1}$ derived by Kenney et al. (1993), assuming that the corotation radius is located at the end of the bar. In figure 9 we show the radial velocity of a pattern speed of 77 km s$^{-1}$ kpc$^{-1}$ (a line). It is apparent that there are velocity components which are smaller than the radial velocity of the pattern. This means that the components rotate slower than the pattern, being inconsistent with the assumption that the corotation radius is located at the end of the bar. Figure 11 shows the P--V diagram along the ridge of the upper side in figure 2 (along the line of $X = -11''$). The distribution of CO emission is not symmetric about $Y = 0''$; in the northen (upper in the disgram) side where the line of $X = -11''$ traces the molecular ridge, the velocity width (e.g., 82 km s$^{-1}$ at $Y = 22''$) is larger than that in the southern (lower in the diagram) side (e.g., 36 km s$^{-1}$ at $Y = -22''$). This asymmetry is due to the velocity change at the ridge. The dashed line represents the radial velocity of the pattern expected from $\Omega_{\rm p}$ = 77 km s$^{-1}$ kpc$^{-1}$. The figure again indicates that there are components slower than the radial velocity of the pattern. These results mean that the pattern speed derived assuming that the corotation radius is located at the end of the bar is too large. On the other hand, the radial velocity expected from the upper limit of 31 km s$^{-1}$ kpc$^{-1}$ (solid line in figure 11) traces the lower velocity edge of the contours which looks like rigid rotation because of the reason mentioned above. 


\begin{fv}{11}
{0cm}
{Position--velocity diagram along the line of $X = -11''$. The lowest contour and contour interval are $T_{\rm MB}$ = 40 mK and 20 mK, respectively. The solid and dashed lines indicate the radial velocities of the patterns with pattern speeds of $\Omega_{\rm p}$ = 31 km s$^{-1}$ kpc$^{-1}$ and 77 km s$^{-1}$ kpc$^{-1}$, respectively. The systemic velocity of 1535 km s$^{-1}$ is also given by a thin line. The arrow indicates the bar length.}
\end{fv}

The misalignment of the major axis of the bar and the line of nodes of the galaxy makes the upper limit of the pattern speed larger. Since, however, the misalignment is smaller than $6^{\circ}$ (Kenney et al. 1993), the difference is smaller than 10 km s$^{-1}$ kpc$^{-1}$, even if we assume that the velocity  along the ridge is 200 km s$^{-1}$ (i.e., the rotation velocity at the radius of 22$''$) and that the molecular gas moves along the bar. Therefore, we conclude that the upper limit of the pattern speed in NGC 3504 is 31 + 10 km s$^{-1}$ kpc$^{-1}$ = 41 km s$^{-1}$ kpc$^{-1}$ incorporating a probable misalignment between the bar major axis and the line of nodes of the galaxy. 
 
\subsection
{Locations of Resonances}
We plotted the radial behavior of the rotation curve, $\Omega$, $\Omega - \kappa/2$ and $\Omega - \kappa/4$ in figure 12. In most galaxies whose pattern speed of the bar is derived assuming the location of a resonance, the corotation radius, $R_{\rm c}$, is within twice of the radius of the bar end, $R_{\rm bar}$ (Elmegreen 1996). However, from the upper limit of the pattern speed we derived  (a dashed line at $\Omega_{\rm p}$ = 41 km s$^{-1}$ kpc$^{-1}$ in figure 12), $R_{\rm c} > 66''$ and the ratio $R_{\rm c}/R_{\rm bar}$ is larger than 2 in NGC3504 ($R_{\rm bar}  = 32''$ : Kenney et al.\ 1993).

If we assume that the molecular gas moves along the bar after shock at the bar, the upper limit derived here is close to the pattern speed. Thus, we suppose the upper limit to be the pattern speed.
In that case, an inner inner Lindblad resonance (IILR) and an outer inner Lindblad resonance (OILR) are located at $R \sim 1''$ and $16''$, respectively. It is interesting that the radius of OILR roughly coincides with the length of the elongated structure in the center, which is tilted with respect to the bar (see section 3). Furthermore, the very small radius of the IILR corresponds to the radius of the star-forming ring ($R = 2''$) found by Planesas et al. (1997). These results are consistent with the starburst scenario in barred spirals which have two ILRs (Combes, Gerin 1985). In this scenario, a very prominent nuclear ring is formed inside the IILR, while a ring formed at the OILR is not prominent. As a result, intense star formation occurs near the IILR. In such galaxies, since the orientation of the elongated orbit of gas changes by $90^{\circ}$ at each resonance, the orbit is perpendicular to the bar between the two ILRs (Combes 1988). 


\begin{fv}{12}
{0cm}
{Radial distributions of the rotation curve (dotted line), $\Omega$, $\Omega-\kappa/2$ and $\Omega-\kappa/4$ in NGC 3504. The dashed line indicates the pattern speed of 41 km s$^{-1}$ kpc$^{-1}$.}
\end{fv}

Although the curve of $\Omega-\kappa/4$ in figure 12 seems to reach the pattern speed between the bar end ($R = 32''$) and the outer ring ($R = 64''$), the uncertainty of the rotation curve is too large to determine the location of the ultraharmonic resonance ($\Omega_{\rm p} = \Omega-\kappa/4$).

\section{Summary}
In this paper, we presented the distribution and kinematics of the molecular gas in the barred spiral galaxy, NGC 3504. The results are summarized as follows:

(1) The molecular gas is highly concentrated in the central region of NGC 3504. The molecular gas is also elongated along the leading edges of the bar. The molecular ridges trace H\,{\footnotesize II} regions along the bar.

(2) There is a large velocity gradient of 50 km s$^{-1}$ at the molecular ridge, which corresponds to 130 km s$^{-1}$ in the galactic disk. The large velocity gradient indicates that the velocity component perpendicular to the bar decreases abruptly at the ridge. The rate of deceleration of 84 km s$^{-1}$ kpc$^{-1}$ is comparable with that in the arms of the grand-design spiral galaxy M 51. 

(3) Since the position angle of the bar and the line of nodes of NGC 3504 are almost the same, we can derive an upper limit to the pattern speed of the bar from the radial velocity at the molecular ridge where molecular gas moves nearly along the bar. The upper limit of the pattern speed is 41 km s$^{-1}$ kpc$^{-1}$, which is much lower than that derived assuming that the location of the corotation radius is the bar end (77 km s$^{-1}$ kpc$^{-1}$). 

(4) The derived pattern speed of the bar shows that there are two ILRs in NGC 3504. It is suggested that the intense star formation in NGC 3504 is occurring at the IILR, as suggested by numerical simulations.
\par
\vspace{1pc}\par
\section*{References}

\re
Athanassoula E.\ 1992, MNRAS 259, 345
\re
Canzian B.\ 1993, PASP 105, 661
\re
Combes F.\ 1988, in Galactic and Extragalactic Star Formation, ed R.E.\ Pudritz, M.\ Fich (Kluwer Academic Publishers, Boston, Dordrecht) p475
\re
Combes F.\ 1996, in Barred Galaxies, ed R.\ Buta, D.A.\ Crocker, B.G.\ Elmegreen, ASP Conf.\ Ser.\ 91, p286
\re
Combes F., Gerin M.\ 1985, A\&A 150, 327
\re
Combes F., Elmegreen B.G.\ 1993, A\&A 271, 391
\re
Condon J.J., Helou G., Sanders D.B., Soifer B.T.\ 1990, ApJS 73, 359
\re
de Vaucouleurs G., de Vaucouleurs A., Corwin H.G., Jr., Buta R.J., Paturel G., 
Fouque P.\ 1991, Third Reference Catalog of Bright Galaxies (Springer-Verlag, 
New York) (RC3)
\re
Elmegreen B.G.\ 1996, in Barred Galaxies, ed R.\ Buta, D.A.\ 
Crocker, B.G.\ Elmegreen, ASP Conf.\ Ser.\ 91, p197
\re
Grosb{\o}l P.J.\ 1985, A\&AS 60, 261
\re
Handa T., Ishizuki S., Kawabe R.\ 1994, in Astronomy with Millimeter and Submillimater Wave Interferometry, ed M.\ Ishiguro, Wm.\ J.\ Welch, ASP Conf.\ Ser.\ 59, p341
\re
Handa T., Nakai N., Sofue Y., Hayashi M., Fujimoto M.\ 1990, PASJ 42, 1
\re
H\"uttemeister S., Aalto S., Wall W.F.\ 1999, A\&A 346, 45
\re
Kenney J.D.P., Carlstrom J.E., Young J.S.\ 1993, ApJ 418, 687
\re
Kuno N., Nakai N.\ 1997, PASJ 49, 279
\re
Laine S., Kenney J.D.P., Yun M.S., Gottesman S.T.\ 1999, ApJ 511, 709
\re
Nakai N.\ 1992, PASJ 44, L27
\re
Nishiyama K., Nakai N.\ 2000, PASJ submitted 
\re
Peterson C.J.\ 1982, PASP 94, 409
\re
Planesas P., Colina L., Prez-Olea D.\ 1997, A\&A 325, 81
\re
Regan M.W., Sheth K., Vogel S.N.\ 1999, ApJ 526, 97
\re
Sandage A.R.\ 1961, The Hubble Atlas of Galaxies (Carnegie Institution of Washington, Washington DC)
\re
Sempere M.J., Garcia-Burillo S.\ 1997, A\&A 325, 769
\re
Sempere M.J., Garcia-Burillo S., Combes F., Knapen J.H.\ 1995, A\&A 296, 45
\re
Sorai K., Nakai N., Kuno N., Nishiyama K., Hasegawa T.\ 2000, PASJ 52,
\re
Strong A.W., Bloemen J.B.G.M., Dame T.M., Grenier I.A., Hermsen W., Lebrun F., Nyman L.-\AA., Pollock A.M.T., Thaddeus P.\ 1988, A\&A 207, 1
\re
Sunada K., Noguchi T., Tsuboi M., Inatani J.\ 1995, in Multi-feed Systems for Radio 
Telescopes, ed D.T.\ Emerson, J.M.\ Payne, ASP Conf.\ Ser.\ 75, p230
\re
Tremaine S., Weinberg M.D.\ 1984, ApJ 282, L5
\re
Ulich B.L., Haas R.W.\ 1976, ApJS 30, 247
\re
van Moorsel G.A. 1983, A\&AS 54, 1

\label{last}
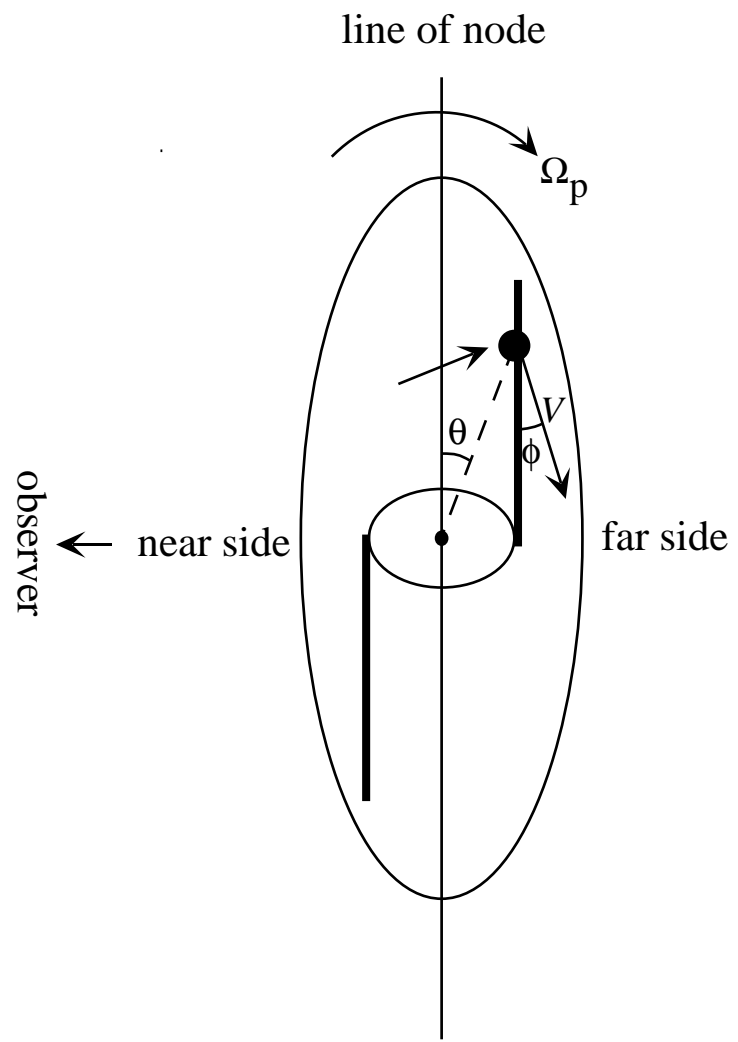
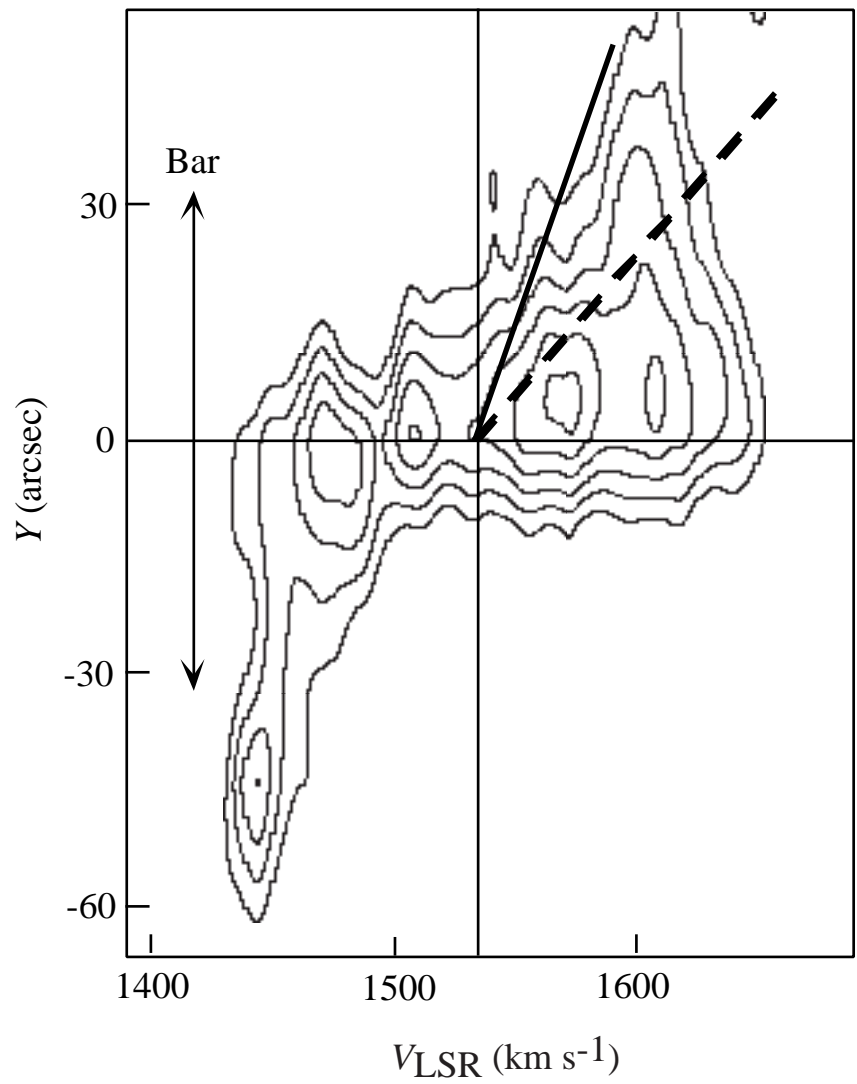
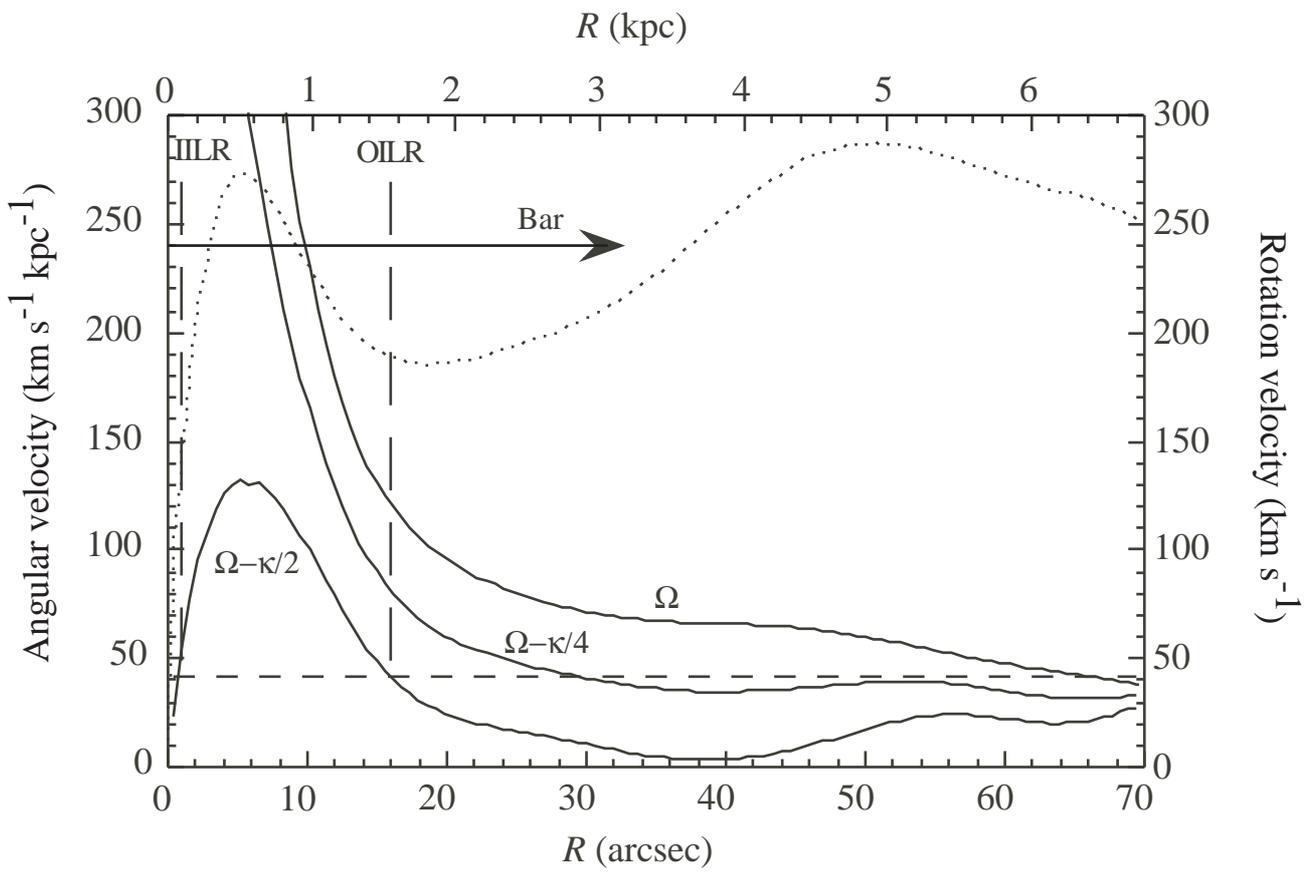
\end{document}